\documentclass{article}
\usepackage[main, final, nonatbib]{neurips_2025}

\usepackage[utf8]{inputenc}
\usepackage[T1]{fontenc}
\usepackage{hyperref}
\usepackage{url}
\usepackage{booktabs}
\usepackage{amsfonts}
\usepackage{nicefrac}
\usepackage{microtype}
\usepackage{xcolor}
\usepackage[numbers]{natbib}

\title{Dataset of Yul Contracts to Support Solidity Compiler Research}

\author{
  Krzysztof Fonał \\
  Faculty of Electronics, Photonics and Microsystems\\
  Wrocław University of Science and Technology\\
  Wrocław, Poland \\
  \texttt{krzysztof.fonal@pwr.edu.pl}
}

\begin{document}

\maketitle

\begin{abstract}
The YulCode dataset presents a comprehensive collection of 348,840 Yul-based smart contract instances, comprising approximately 135,013 unique contracts. These contracts were generated through the compilation of Solidity source files that have been deployed on the Ethereum mainnet, making the dataset directly representative of real-world decentralized applications. YulCode provides a rich foundation for a variety of research and development tasks, including but not limited to machine learning applications, formal verification, optimization analysis, and software engineering tool evaluation in the context of low-level smart contract code. To the best of our knowledge at the time of writing, YulCode is the first and only publicly available dataset that focuses specifically on Yul, an intermediate language designed for the Ethereum Virtual Machine (EVM). As such, it fills a critical gap in the current ecosystem of smart contract datasets and opens new avenues for research and tooling aimed at low-level contract analysis and generation.
\end{abstract}

\section{Introduction}
Smart contracts are self-executing programs that run on top of decentralized ledgers, most notably blockchain platforms such as Ethereum~\cite{ethereum}. Their use cases span a wide array of domains, including finance, governance, gaming, and digital art. Among the various programming languages for smart contract development, \textbf{Solidity} remains the most widely adopted~\cite{solidity}.

As artificial intelligence (AI) continues to influence nearly every field, software engineering is emerging as one of its most promising applications. However, the success of AI methods in software engineering heavily relies on the availability of high-quality datasets. While there exist numerous datasets for conventional programming languages (e.g., Python, Java, JavaScript), the blockchain domain suffers from a relative lack of such resources. 

Several efforts have addressed Solidity datasets~\cite{bigquery, fiesta, disl}, with DISL~\cite{disl} being among the most comprehensive containing over 400,000 Solidity contracts as of January 15, 2024. These datasets are invaluable for tasks such as vulnerability detection, code summarization, and automated testing.

Despite these advances, a crucial layer of the smart contract compilation pipeline remains underserved: the intermediate representation known as \textbf{Yul}~\cite{yulref}. The Solidity compiler (\texttt{solc}) translates high-level Solidity code into Yul before generating the final EVM bytecode. Most of the compiler’s optimization passes are performed at the Yul level, making it a critical stage for low-level reasoning, performance tuning, and formal verification. In performance-critical scenarios, some developers even write contracts directly in Yul to gain finer control over the resulting bytecode.

Currently, there is a complete lack of publicly available datasets targeting Yul code, which limits research and tooling development focused on compiler internals, bytecode optimization, and intermediate representations. This work addresses this gap by introducing the \textbf{YulCode} dataset, a collection of nearly 350,000 smart contract instances expressed in Yul. This dataset can be used to support a wide range of downstream research tasks, with a particularly promising direction being the study of the \textit{phase-ordering problem} in the Solidity compiler—an optimization challenge due to the hardcoded sequence of transformation passes on Yul code.

\section{Dataset Collection Process}

To construct the YulCode dataset, I used the DISL dataset of Solidity smart contracts as the foundation~\cite{disl}. The creation pipeline for YulCode followed these main steps:

\begin{enumerate}
    \item Traverse each Solidity contract available in the DISL dataset.
    \item Extract the Solidity compiler version specified in the contract.
    \item If the compiler version is \texttt{0.8.10} or later—which is the first version that supports the generation of Yul intermediate representation (IR)—the contract is compiled using the corresponding version of the \texttt{solc} compiler.
\end{enumerate}

In practice, many contracts required an additional setup step to resolve external dependencies. A common pattern observed was the use of widely adopted libraries such as OpenZeppelin. For such cases, it was often sufficient to set the \texttt{--include-path} option to point to the \texttt{node\_modules} directory containing these shared libraries.

However, not all contracts followed the same dependency structure. Some contracts referenced local libraries residing in the same directory as the main contract. For those, all necessary files were programmatically copied into a single directory to allow successful compilation.

As a result of this process, for each Solidity contract, we obtained at least two corresponding Yul representations:
\begin{itemize}
    \item The initialization (constructor) Yul code, executed at deployment time.
    \item The runtime Yul code, representing the deployed contract logic.
\end{itemize}

In cases where a Solidity contract includes or depends on other contracts (e.g., libraries or interfaces), multiple additional Yul fragments may be generated.

By the end, a total of 348,840 Yul contracts were collected. While there are many duplicates due to repeated use of well-known libraries, I chose not to remove them, as slight variations can exist across different versions of the Solidity compiler (solc). However, I performed an analysis of the duplicated contracts, and after filtering out common libraries such as Ownable, SafeMath, and ERC20, the number of unique, non-library contracts was found to be 135,013.

The code used to perform the dataset conversion and compilation process is publicly available at: 

\url{https://github.com/krzysiekfonal/solc-optimizer/blob/main/yul_source_generator.py} .

\section{Dataset Content}

The \textbf{YulCode} dataset consists of a comprehensive collection of Yul intermediate representation (IR) files generated from Solidity smart contracts. Each data entry includes several key fields that provide detailed metadata about the contract and its compilation context. The dataset fields are as follows:

\begin{itemize}
    \item \textbf{contract\_name} – The name of the smart contract to which the Yul file corresponds. This is typically extracted from the Solidity source file.
    
    \item \textbf{contract\_address} – The Ethereum address at which the contract has been deployed. This allows for correlation between on-chain deployed contracts and their Yul representations.
    
    \item \textbf{solc\_version} – The specific version of the Solidity compiler (\texttt{solc}) used to compile the contract. This is important because the Yul output can vary between compiler versions.
    
    \item \textbf{sol\_filepath} – The full or relative file path to the original Solidity (\texttt{.sol}) source file used during compilation.
    
    \item \textbf{yul\_filepath} – The file path to the corresponding Yul output file generated by the compiler. This allows for traceability between the source and intermediate representation.
    
    \item \textbf{source\_code} – The Yul code itself, produced by the Solidity compiler as an intermediate representation. This code is useful for low-level analysis, optimization research, and compiler studies.
\end{itemize}

The dataset is available publically on HugginFace at:

\url{https://huggingface.co/datasets/KrzysztofFonal/yulcode}.

\section{Applications}

The \textbf{YulCode} dataset provides a valuable resource for a variety of research and development tasks related to compiler design, smart contract analysis, and machine learning. Some notable applications include:

\begin{itemize}
    \item \textbf{Compiler Phase-Ordering Exploration} – Yul, as an intermediate representation, allows researchers to experiment with different phase-ordering strategies within the Solidity compiler pipeline. This dataset can serve as input for evaluating how various combinations of compiler optimization passes affect the performance, size, and gas efficiency of the resulting EVM bytecode.

    \item \textbf{Training Data for Neural Models} – The dataset provides a large and diverse corpus of Yul code suitable for training neural networks, such as transformers or graph neural networks. These models can be applied to tasks like code generation, optimization prediction, decompilation, vulnerability detection, or automated compiler pass sequencing.

    \item \textbf{IR-Level Vulnerability Analysis} – Since Yul sits closer to the EVM than Solidity, it offers an ideal abstraction level for analyzing low-level contract behaviors. The dataset can support the development of static or symbolic analysis tools focused on detecting security issues at the IR level.

    \item \textbf{Cross-Version Consistency Testing} – Because the dataset includes metadata about the Solidity compiler version used to generate each Yul file, it enables studies on how the IR changes across compiler versions and whether those changes are consistent, correct, or introduce regressions.

    \item \textbf{EVM Bytecode Generation Benchmarking} – By compiling Yul files into bytecode, developers can assess the performance and correctness of different Yul-to-EVM compilers or backend code generators, which is useful in both academic research and production compiler engineering.

    \item \textbf{Smart Contract Reverse Engineering} – The dataset can support efforts to reverse engineer deployed contracts by correlating Yul IR with on-chain bytecode and original source files, helping to understand obfuscation strategies or undocumented behaviors.
\end{itemize}

In summary, the \textbf{YulCode} dataset opens up numerous possibilities for advancing smart contract toolchains, improving compiler technologies, and enabling data-driven approaches to program analysis.

\section{Conclusion}

The \textbf{YulCode} dataset presents a large-scale, structured collection of Yul intermediate representations derived from real-world Solidity smart contracts. By capturing detailed metadata such as contract names, deployment addresses, compiler versions, and file mappings, the dataset provides a rich foundation for a wide range of research and development efforts.

From investigating compiler phase-ordering strategies to training neural networks for IR-level tasks, YulCode facilitates exploration at the intersection of compiler design, smart contract security, and machine learning. Its inclusion of both raw Yul code and contextual information enables reproducibility, benchmarking, and comparative studies across Solidity compiler versions and optimization strategies.

I believe this dataset will serve as a valuable resource for the blockchain research community and compiler engineers, fostering advancements in automated optimization, secure compilation, and smart contract analysis. Future work may include expanding the dataset to cover additional compiler backends, integrating runtime profiling data, or curating labeled subsets for supervised learning tasks.

\end{document}